\documentclass{JHEP3}

\usepackage{epsfig}
\usepackage{amsbsy}
\usepackage{varioref}
\usepackage{pifont}

\title{Higgsless Electroweak Symmetry Breaking from Theory Space}

\author{Roshan Foadi, Shrihari Gopalakrishna, Carl Schmidt\\
Department of Physics and Astronomy, Michigan State University\\
East Lansing, MI 48824, USA\\
	E-mail: \email{foadi@pa.msu.edu}, 
\email{shri@pa.msu.edu},
\email{schmidt@pa.msu.edu}}

\abstract{
We investigate unitarity of $W^+W^-$ scattering in the context of
theory space models of the form $U(1)\times {[SU(2)]}^N\times SU(2)_{N+1}$,
which are broken down to $U(1)_{EM}$ by non-linear $\Sigma$ fields, without
the presence of a physical Higgs Boson.  By allowing the couplings of the
$U(1)$ and the final $SU(2)_{N+1}$ to vary, we can fit the $W$ and $Z$
masses, and we find that the coefficient of the term in the amplitude 
that grows as $E^2/m_W^2$ at high energies is suppressed by a factor
of $(N+1)^{-2}$.  In the $N+1\rightarrow\infty$ limit the model becomes
a 5-dimensional $SU(2)$ gauge theory defined on an interval, where
boundary terms at the two ends of the interval break the $SU(2)$ down
to $U(1)_{EM}$.  These boundary terms also modify the Kaluza-Klein (KK)
mass spectrum, so that the lightest KK states can be identified as
the $W$ and $Z$ bosons.  The $T$ parameter, which measures custodial
symmetry breaking, is naturally small in these models.  Depending on how 
matter fields are included, the strongest experimental constraints come 
from precision electroweak limits on the $S$ parameter.
}

\preprint{{~MSUHEP--031216}
{~hep-ph/0312324}}

\begin{document}


\section{Introduction}
\label{sec:Intro}

Despite the successes of the standard model of particle
physics, we know that it is incomplete without some, as yet, 
undiscovered particles and interactions.  The simplest argument for
this is to consider the scattering of longitudinal vector
bosons at high energies.  The scattering amplitude for
$W_L^+W_L^-\rightarrow W_L^+W_L^-$ grows as $E^2/m_W^2$
at high energies, and therefore violates unitarity at some
scale.  This is the main motivation for
the existence of one or more Higgs scalars, which naturally
fix this problem by providing new contributions to the
scattering amplitude~\cite{WWscattering}.  The perturbative
breaking of the $SU(2)\times U(1)$
symmetry by Higgs vacuum expectation values (vevs) is intimately
related with this unitarity restoration.  An alternative path to 
symmetry breaking is some new strong dynamics, such as 
technicolor~\cite{technicolor}.
In these types of models the unitarity is restored through the
exchange of composite states, such as techni-rho mesons.  

Recently, a new mechanism for gauge-symmetry breaking has been 
suggested~\cite{edbreaking,Csaki:2003dt}.
This involves the embedding of the 4-dimensional theory in some
higher-dimensional model, where the gauge symmetry is exact in
the bulk of the extra dimensions, but it is explicitly broken on
the boundaries.  The breaking can either be due to orbifolding the
extra dimensions or to simply imposing conditions on the fields at
the boundary, which 
distinguish between subsets of the full gauge group~\cite{Csaki:2003dt}.  
Despite the explicit symmetry breaking in the higher-dimensional theory,
it appears soft in the effective 4-dimensional theory, in the sense
that the unitarity of scattering amplitudes is 
maintained~\cite{sekhar}.\footnote{
Of course, at sufficiently high energies, unitarity problems will
reappear, due to the fact that the higher-dimensional theory is
not renormalizable.  
Although this must be addressed in a complete
self-consistent theory, it is beyond the scope of the ideas considered
in this paper.}  In this case the unitarity of the longitudinal
vector boson scattering amplitude is restored via the exchange of
Kaluza-Klein (KK) excitations of the gauge bosons.

Models of this sort have been used to explain the breaking
of a grand unified gauge group down to the standard model 
$SU(3)\times SU(2)\times U(1)$. (See \cite{edbreaking}, for example.)  Attempts have also been made
to use this mechanism to break the electroweak gauge group
$SU(2)\times U(1)$ down to electromagnetism 
$U(1)_{EM}$~\cite{csaki,nomura,barbieri}
without the necessity for a Higgs scalar.
In this paper we use the technique of deconstruction of the
extra dimension~\cite{deconstruction} as an aid to addressing this
second idea.  In deconstruction of a five-dimensional gauge theory
the fifth dimension is discretized, and the gauge fields $A^{\mu}$
at each position in the extra dimension become independent gauge fields
of a product gauge group in four dimensions.  The gauge fields that point 
along the fifth dimension, $A^5$, are reinterpreted as the Goldstone-boson 
fields of a 
non-linear sigma model, which break the gauge groups at neighboring sites
of the discretized extra dimension down to the diagonal.
We shall see that investigating deconstructed models 
offers a new insight into this mechanism of symmetry breaking 
and leads us to a new model, which is arguably simpler than those 
proposed previously, even after the continuum limit is recovered.

A useful example of deconstruction of a theory with symmetry-breaking
boundary conditions is the $SU(2)$ model of Ref.~\cite{Csaki:2003dt},
defined on a fifth-dimensional interval, $0\le y\le \pi R$, where
the $SU(2)$ group is broken down to $U(1)$ at $y=0$ by boundary conditions.
The deconstruction of this model is a $U(1)\times {[SU(2)]}^{N+1}$
gauge theory with sigma model fields that break the $SU(2)\times SU(2)$
at neighboring sites down to the diagonal.  (The $U(1)$ at the first
site is realized by gauging the $U(1)$ subgroup of a global $SU(2)$.) 
An interesting observation is that the $N=0$ case of this deconstructed
theory just corresponds to the electroweak gauge group, except that in 
the deconstructed model the $U(1)$ and $SU(2)$ couplings are identical.  
If we allow the couplings to be different, we have the effective field 
theory for a standard model in which the electroweak symmetry is broken 
strongly, such as technicolor.  Thus, for general $N$, one might consider
whether the good unitarity properties of the deconstructed theory can be 
maintained, while relaxing the conditions on the gauge couplings of the product
groups and on the vevs of the sigma model fields.  In the remainder of
this paper, we follow this line of reasoning and see where it leads.

In section \ref{sec:model1} we consider the simplest extension
of the (Higgsless) standard model of this type, which is 
$U(1)\times SU(2)_1\times SU(2)_2$.  Allowing the three
couplings and the two vacuum expectation values (which break the
group down to $U(1)_{EM}$) to be independent,
we can fit the lightest charged and neutral vector bosons
to the $W$ and $Z$ masses and the remnant gauge coupling to the
electromagnetic coupling $e$.  We then investigate the question
of unitarity of the $W^+_LW^-_L\rightarrow W^+_LW^-_L$ scattering
amplitude as a function of the two remaining parameters, which can
be taken to be the masses of the $W^\prime$ and $Z^\prime$ which
arise in this theory.  We find that the unitarity violation is postponed
to the greatest extent when the two vevs of the sigma fields are
roughly equal.  When this condition is satisfied, we find that 
$m_{W^\prime}$ and $m_{Z^\prime}$ are approximately proportional to
the coupling of the central $SU(2)_1$, while the $W$ and $Z$ masses
and couplings are essentially given by those of the $U(1)$ and $SU(2)_2$
on the ends of the chain.

Motivated by those results, we then consider in section 
\ref{sec:model2} the generalization
to $U(1)\times {[SU(2)]}^N\times SU(2)_{N+1}$, where all of the
vevs of the sigma fields are taken identical, and all of the
couplings except for those of the $U(1)$ and the $SU(2)_{N+1}$ on
the end are taken identical.  We find that in the limit of 
$N+1\rightarrow\infty$ the same good unitarity behavior of the
extra-dimensional theory is recovered.  In section \ref{sec:edim} 
we consider the $N+1\rightarrow\infty$ limit further and 
examine the five-dimensional theory that is obtained.  
In section \ref{sec:fermions} we investigate the question of the
inclusion of fermions in these models, and analyze 
the direct and indirect constraints from experiment.
Finally, in section \ref{sec:conclusions} we give our conclusions.
We also have included two appendices which give more details
of the solutions to the $N=1$ and general $N$ 
models considered in sections \ref{sec:model1} and \ref{sec:model2}, 
respectively.

\section{$U(1)\times SU(2)_1\times SU(2)_2$ Model}
\label{sec:model1}

\EPSFIGURE[t]{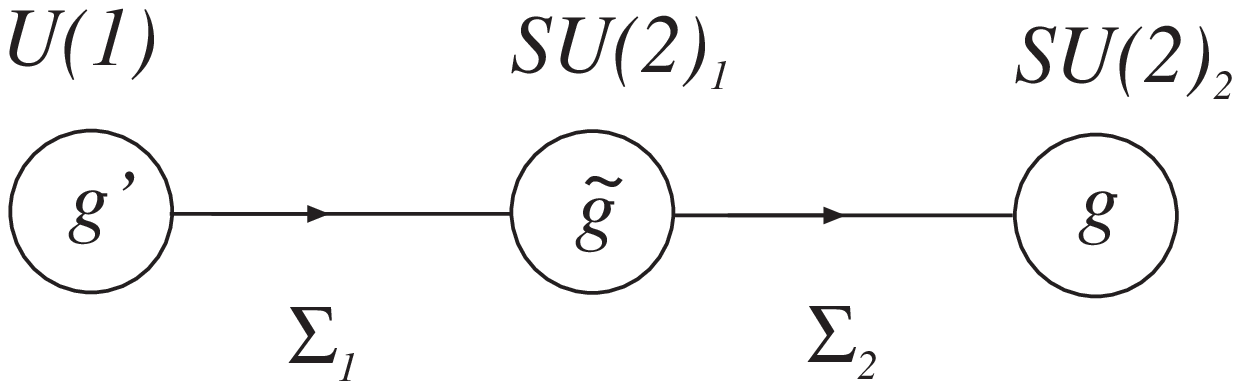,width=0.6\textwidth}
{Moose diagram for $U(1)\times SU(2)_1\times SU(2)_2$ model.
\label{fig:Moose1}}

We begin with the theory space model with the group structure
$U(1)\times SU(2)_1\times SU(2)_2$ and defined by the moose
diagram~\cite{mooses} of Fig.~\ref{fig:Moose1}, where the $U(1)$ is
treated as the $U(1)$ subgroup of a global $SU(2)$ group.  
The nonlinear sigma model fields,
\begin{equation}
\Sigma_1\ =\ e^{2i\pi_1^aT^a/f_1}\ ,\qquad\ \Sigma_2\ =\ e^{2i\pi_2^aT^a/f_2}\ ,
\label{eq:sigmas}
\end{equation}
consist of two $SU(2)$ triplets, which are coupled to the gauge fields
by the covariant derivatives
\begin{eqnarray}
D_\mu\Sigma_1&=& \partial_\mu\Sigma_1-ig^\prime T^3B_\mu\Sigma_1+i\tilde{g}\Sigma_1T^aW^a_{1\mu}\ ,\nonumber\\
D_\mu\Sigma_2&=& \partial_\mu\Sigma_2-i\tilde{g} T^aW^a_{1\mu}\Sigma_2+ig
\Sigma_2T^aW^a_{2\mu}\ .
\label{eq:covdiv}
\end{eqnarray}
The Lagrangian density for the relevant fields is 
\begin{eqnarray}
{\cal L}&=& -{1\over4}B^{\mu\nu}B_{\mu\nu}-{1\over4}W_1^{a\,\mu\nu}W^a_{1\,\mu\nu}
-{1\over4}W_2^{a\,\mu\nu}W^a_{2\,\mu\nu}\nonumber\\
&&
+{f_1^2\over4}{\rm tr}\Big[D^\mu\Sigma_1(D_\mu\Sigma_1)^\dagger\Big]
+{f_2^2\over4}{\rm tr}\Big[D^\mu\Sigma_2(D_\mu\Sigma_2)^\dagger\Big]\ .
\label{eq:lagrange1}
\end{eqnarray}
After the sigma fields acquire a vacuum expectation value, 
$\langle\Sigma_i\rangle=1$, their only affect in unitary gauge 
is to give mass to the $W$'s and $Z$'s.

In this model there are five independent parameters: $g^\prime,\,
\tilde{g},\,g,\,f_1,\,f_2$.  Since we want to recover the standard
model at low energy, we can exchange three of these for the 
electromagnetic coupling $e$, and the $W$ and $Z$ boson masses,
$m_W$ and $m_Z$.  (Note that we are only working at tree level throughout
this paper.)  It is convenient to take the masses of the heavy $W^\prime$
and $Z^\prime$ bosons, $m_{W^\prime}$ and $m_{Z^\prime}$, as the remaining
independent parameters.  The charged gauge bosons can be expanded in terms of
the mass eigenstates by
\begin{eqnarray}
W_1^\pm&=& a_{11}\,W^{\prime\pm}+a_{12}\,W^{\pm}\nonumber\\
W_2^\pm&=& a_{21}\,W^{\prime\pm}+a_{22}\,W^{\pm}\ .
\label{eq:Ws}
\end{eqnarray}
The neutral bosons can be expanded in terms of mass eigenstates by
\begin{eqnarray}
B&=& b_{00}\,A+b_{01}\,Z^\prime+b_{02}\,Z\nonumber\\
W_1^3&=& b_{10}\,A+b_{11}\,Z^\prime+b_{12}\,Z\nonumber\\
W_2^3&=& b_{20}\,A+b_{21}\,Z^\prime+b_{22}\,Z\ ,
\label{eq:Zs}
\end{eqnarray}
where the photon $A$ is massless.
Formulae for the original parameters $g^\prime,\,
\tilde{g},\,g,\,f_1,\,f_2$, and the mixing matrices, $a_{ij}$, $b_{ij}$
as functions of the independent 
variables $e,\,m_W,\,m_Z,\,m_{W^\prime},\,
m_{Z^\prime}$ can be found in Appendix~\ref{sec:solutions1}.

We now would like to see if it is possible to cancel the bad
high energy behavior in the $W_L^+W_L^-$ scattering amplitude, or at 
least reduce it, for some choice of parameters in our model.
In particular we are concerned with the growth proportional to  
$E^2/m_W^2$ for high energies in the $W_L^+W_L^-\rightarrow W_L^+W_L^-$
amplitude.  The amplitude can be written
\begin{eqnarray}
{\cal A}&=& {1\over m_W^4}\Biggl\{\Bigl(\tilde{g}^2a_{12}^4+g^2a_{22}^4\Bigr)
\Bigl[p^2E^2(-2+6\cos{\theta})+E^4(-1+\cos^2{\theta})\Bigr]\nonumber\\
&&\qquad+
\Bigl[-p^2\cos{\theta}(p^2-3E^2)^2\Bigr]\nonumber\\
&&\qquad\qquad\times
\Biggl[{e^2\over s} 
+ {\bigl(\tilde{g}\,b_{12}\,a_{12}^2+g\,b_{22}\,a_{22}^2\bigr)^2\over s-m_Z^2}
+ {\bigl(\tilde{g}\,b_{11}\,a_{12}^2+g\,b_{21}\,a_{22}^2\bigr)^2\over s-m_{Z^\prime}^2}
\Biggr]
\nonumber\\
&&\qquad-
\Bigl[4E^2\Bigl(p^2+(E^2-2p^2)\cos{\theta}\Bigr)^2
+2p^2(1+\cos{\theta})\Bigl(2E^2-p^2-E^2\cos{\theta})^2
\Bigr]\nonumber\\
&&\qquad\qquad\times
\Biggl[{e^2\over t} 
+ {\bigl(\tilde{g}\,b_{12}\,a_{12}^2+g\,b_{22}\,a_{22}^2\bigr)^2\over t-m_Z^2}
+ {\bigl(\tilde{g}\,b_{11}\,a_{12}^2+g\,b_{21}\,a_{22}^2\bigr)^2\over t-m_{Z^\prime}^2}
\Biggr]\Biggl\}\ ,
\label{eq:amp1}
\end{eqnarray}
where the energy $E$, momentum $p=(E^2-m_W^2)^{1/2}$, and scattering
angle $\theta$ of the $W$'s are in the center of momentum frame, while
$s=4E^2$ and $t=-2p^2(1-\cos{\theta})$ are the standard Mandelstam variables.

\EPSFIGURE[t]{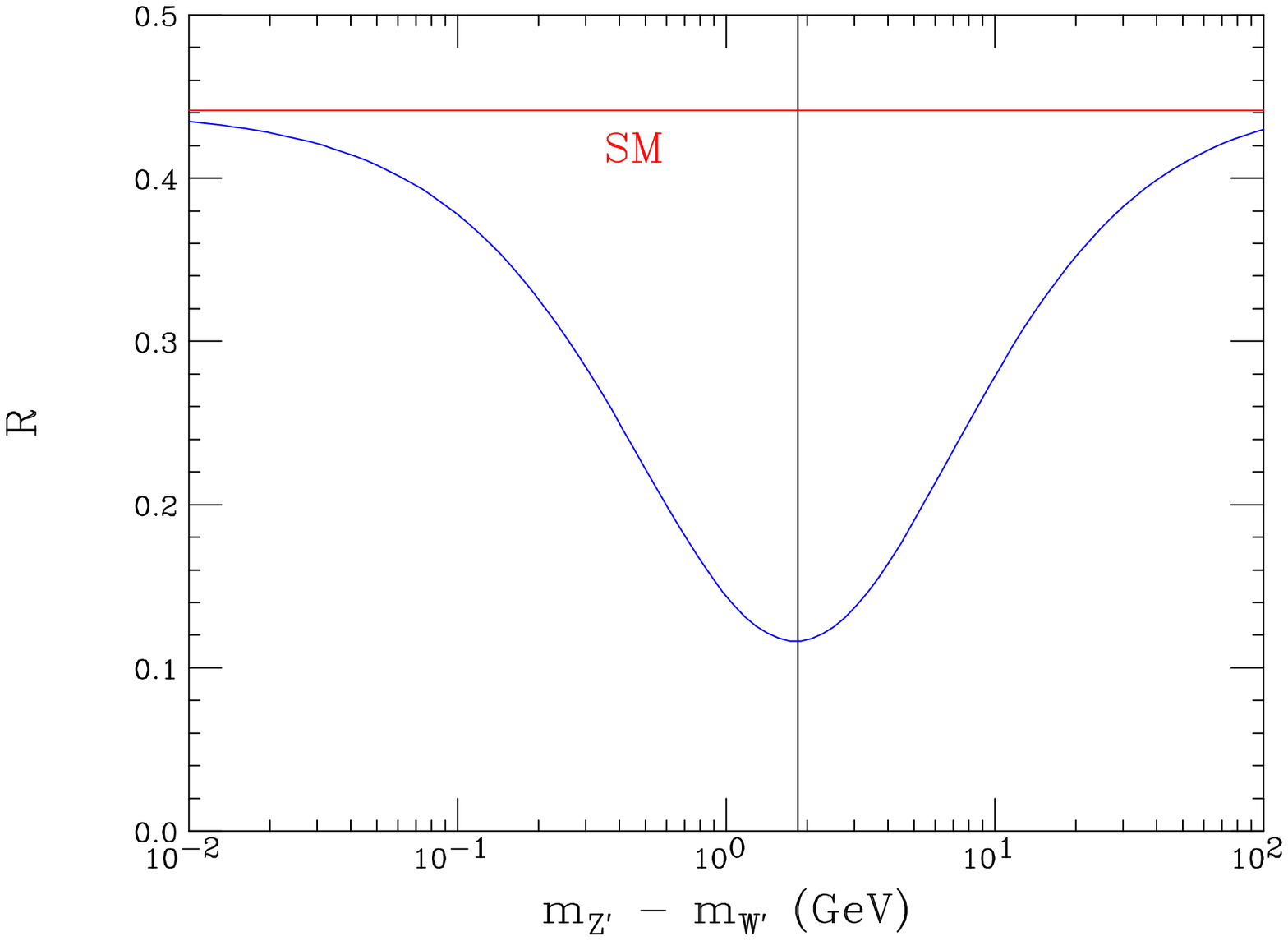,width=0.6\textwidth}
{
The coefficient of the leading $E^2/m_W^2$ term in the $W^+W^-\rightarrow
W^+W^-$ scattering
amplitude in the $U(1)\times SU(2)_1\times SU(2)_2$ model (blue)
as a function of the $Z^\prime$ and $W^\prime$ mass difference, with
$m_{W^\prime}=500$ GeV fixed.  The same quantity in the
standard model without a Higgs boson (red) is also plotted.  The vertical
line indicates the position where  $m_{Z^\prime}^2-m_{W^\prime}^2=m_Z^2-m_W^2$.
\label{fig:R1}}

\EPSFIGURE[t]{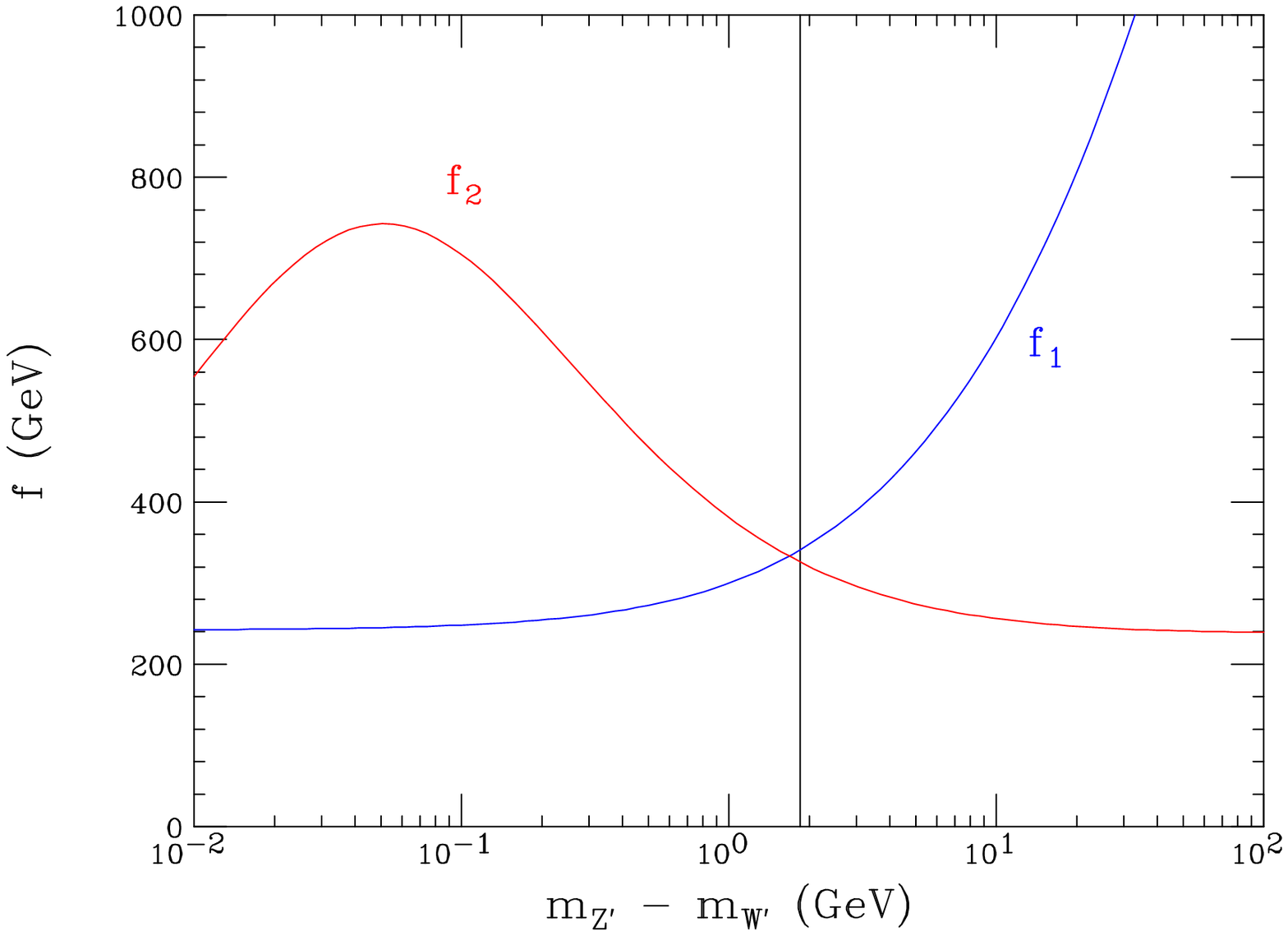,width=0.6\textwidth}
{
The quantities $f_1$ (blue) and $f_2$ (red) 
as a function of the $Z^\prime$ and $W^\prime$ mass difference, with
$m_{W^\prime}=500$ GeV fixed.  The vertical
line indicates the position where  $m_{Z^\prime}^2-m_{W^\prime}^2=m_Z^2-m_W^2$.
\label{fig:f1}}

\EPSFIGURE[t]{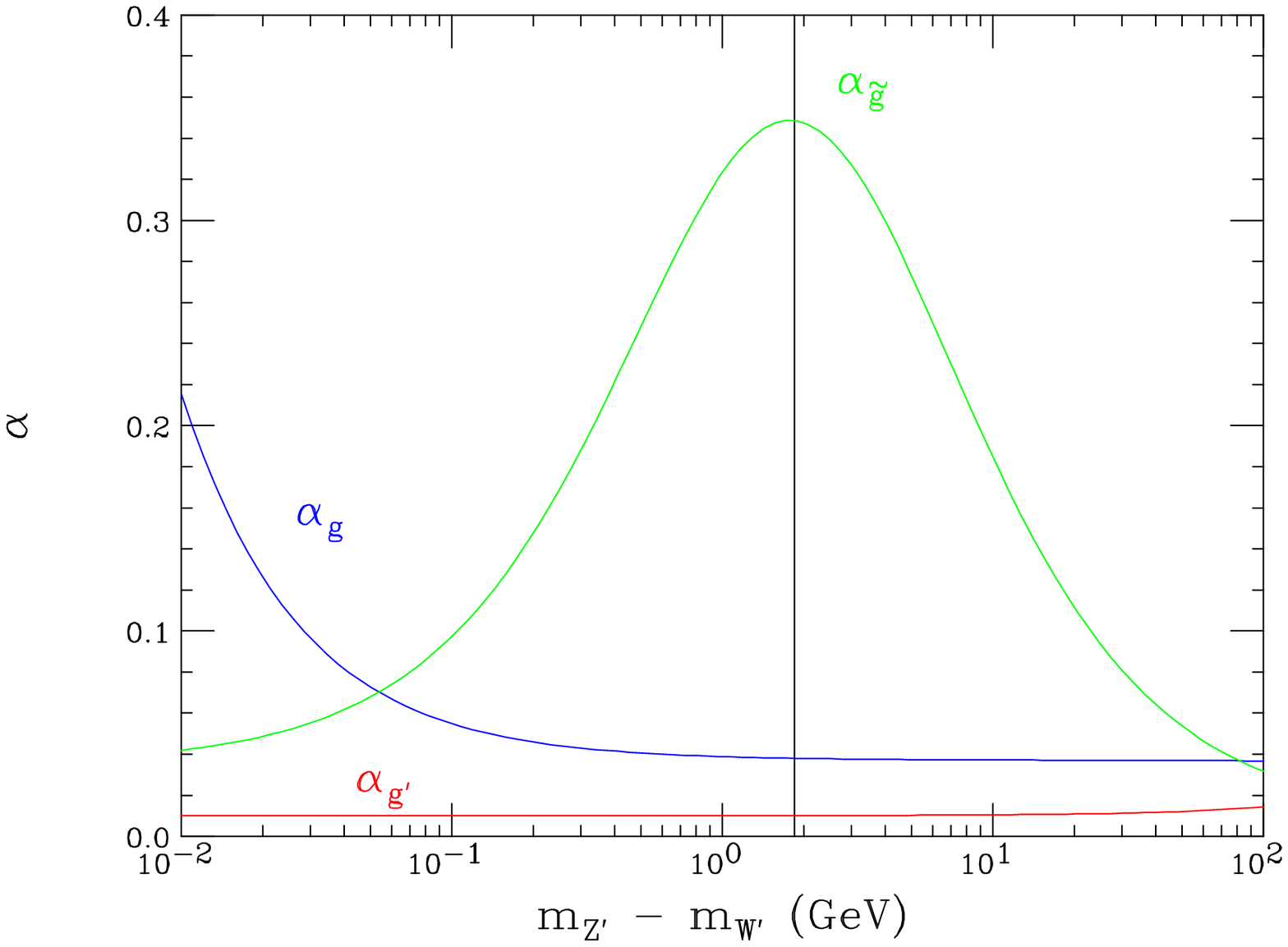,width=0.6\textwidth}
{
The coupling constants $\alpha_g=g^2/4\pi$ (blue),
$\alpha_{g^\prime}=g^{\prime2}/4\pi$ (red), and
$\alpha_{\tilde{g}}={\tilde{g}}^2/4\pi$ (green),
as a function of the $Z^\prime$ and $W^\prime$ mass difference, with
$m_{W^\prime}=500$ GeV fixed.  The vertical
line indicates the position where  $m_{Z^\prime}^2-m_{W^\prime}^2=m_Z^2-m_W^2$.
\label{fig:g1}}

At high energies the amplitude can be expanded in powers of $m_W^2/E^2$.
The coefficient of the leading $E^4/m_W^4$ vanishes due to gauge invariance.
We are then left with
\begin{equation}
{\cal A}\ =\  {E^2\over m_W^2}{(1+\cos{\theta})\over2}R + {\cal O}\Bigl((m_W^2/E^2)^0\Bigr)\ ,
\label{eq:amp1he}
\end{equation}
where
\begin{equation}
R\ =\ 4\Bigl(\tilde{g}^2a_{12}^4+g^2a_{22}^4\Bigr)
-3\Biggl[\bigl(\tilde{g}\,b_{12}\,a_{12}^2+g\,b_{22}\,a_{22}^2\bigr)^2
{m_Z^2\over m_W^2}
+\bigl(\tilde{g}\,b_{11}\,a_{12}^2+g\,b_{21}\,a_{22}^2\bigr)^2
{m_{Z^\prime}^2\over m_W^2}\Biggr]\ .
\end{equation}
Using the formulae in Appendix~\ref{sec:solutions1}, we can treat 
$R\equiv R(m_{W^\prime},m_{Z^\prime})$ as a 
function of $m_{W^\prime}$ and $m_{Z^\prime}$.

In Fig.~\ref{fig:R1} we plot $R$ as a function of the mass difference, 
$m_{Z^\prime}-m_{W^\prime}$, for $m_{W^\prime}=500$ GeV fixed.  
As a comparison we also plot
the same quantity in the standard model without the Higgs boson.  Note that
$R$ is significantly suppressed for $m_{Z^\prime}^2- m_{W^\prime}^2\approx
m_Z^2-m_W^2$.  When this relation holds, the value of $R$ is reduced by
almost precisely a factor of 1/4, a result which does not depend on
the particular value of $m_{W^\prime}$.  This indicates that the unitarity 
violation that occurs
in the standard model without a Higgs boson would be postponed to higher
energy in this model.  We also plot in Fig.~\ref{fig:f1}  the scales $f_1$
and $f_2$, and in Fig.~\ref{fig:g1} the couplings constants 
$\alpha_g=g^2/4\pi$, $\alpha_{g^\prime}=g^{\prime2}/4\pi$, and
$\alpha_{\tilde{g}}={\tilde{g}}^2/4\pi$,
as a function of the $Z^\prime$ and $W^\prime$ mass difference, with
$m_{W^\prime}=500$ GeV fixed.  We note that the relation 
$m_{Z^\prime}^2- m_{W^\prime}^2\approx m_Z^2-m_W^2$ also corresponds
to $f_1\approx f_2$ and $\tilde{g}\gg g,g^\prime$.  In fact when this
relation holds the couplings are given to a good approximation by
$g=e/\sin{\theta_W}$, $g^\prime=e/\cos{\theta_W}$, and 
$\tilde{g}=(m_{W^\prime}/2m_W)g$, up to corrections of order 
$m_W^2/m_{W^\prime}^2$.  (We have used the tree level definition of 
$\cos{\theta_W}=m_W/m_Z$.)
Thus, the $SU(2)_2$ and
the $U(1)$ act approximately like the $SU(2)_L$ and $U(1)_Y$
of the standard model, while the intervening $SU(2)_1$ has the effect
of softening the unitarity violation of the standard model $W_LW_L$
scattering.

\EPSFIGURE[t]{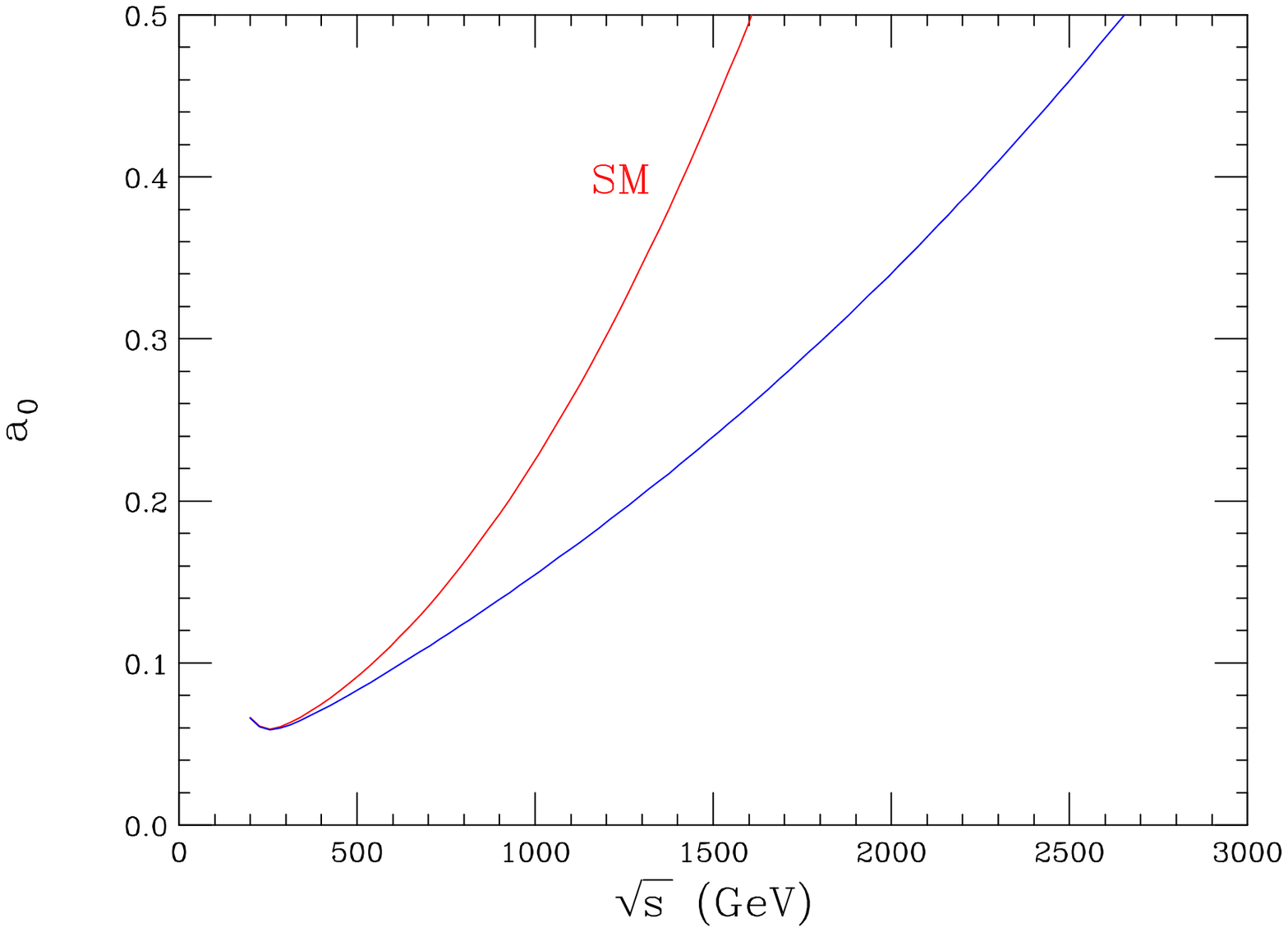,width=0.6\textwidth}
{
The $J=0$ partial wave amplitude as a function of $\sqrt{s}$
for the standard model without a Higgs boson
(red) and the $U(1)\times SU(2)_1\times SU(2)_2$ model
(blue) with $m_{W^\prime}=500$ GeV and
$m_{Z^\prime}^2-m_{W^\prime}^2=m_Z^2-m_W^2$.
\label{fig:a01}}

We can observe the effect of the delayed unitary violation by plotting
the $J=0$ partial wave amplitude,
\begin{equation}
a_0\ =\ {1\over32\pi}\int_{-1}^{1}{\cal A}\, d\cos{\theta}\ ,
\label{eq:j0}
\end{equation}
as a function of $\sqrt{s}=2E$.  This is shown in Fig.~\ref{fig:a01}
for both the standard model without a Higgs boson and in the
$U(1)\times SU(2)_1\times SU(2)_2$ model with $m_{W^\prime}=500$ GeV
and $m_{Z^\prime}^2- m_{W^\prime}^2= m_Z^2-m_W^2$.  
(We have included a photon mass of $m_A=1$ GeV in order to regulate the
 $t$-channel singularity in the integral of Eq.~(\ref{eq:j0}).  This
is inconsequential in the high energy region in which we are interested.)
Since unitarity
requires
$|{\rm Re}\, a_0| <1/2$, we can use this figure to infer that
unitarity violation in this amplitude has been postponed from
a scale of $\sqrt{s}\approx 1.6$ TeV in the standard model without
a Higgs boson to $\sqrt{s}\approx 2.65$ TeV in the
$U(1)\times SU(2)_1\times SU(2)_2$ model with this choice of
parameters.

We have found that the behavior of the 
$W_L^+W_L^-\rightarrow Z_LZ_L$
amplitude to be essentially identical to that for 
$W_L^+W_L^-\rightarrow W_L^+W_L^-$.  In particular the corresponding
value of $R$, the coefficient of the leading $E^2/m_W^2$ term
in that amplitude, is reduced by the same factor of 1/4 when
$m_{Z^\prime}^2- m_{W^\prime}^2\approx m_Z^2-m_W^2$.

\section{$U(1)\times {[SU(2)]}^N\times SU(2)_{N+1}$ Model}
\label{sec:model2}

\EPSFIGURE[t]{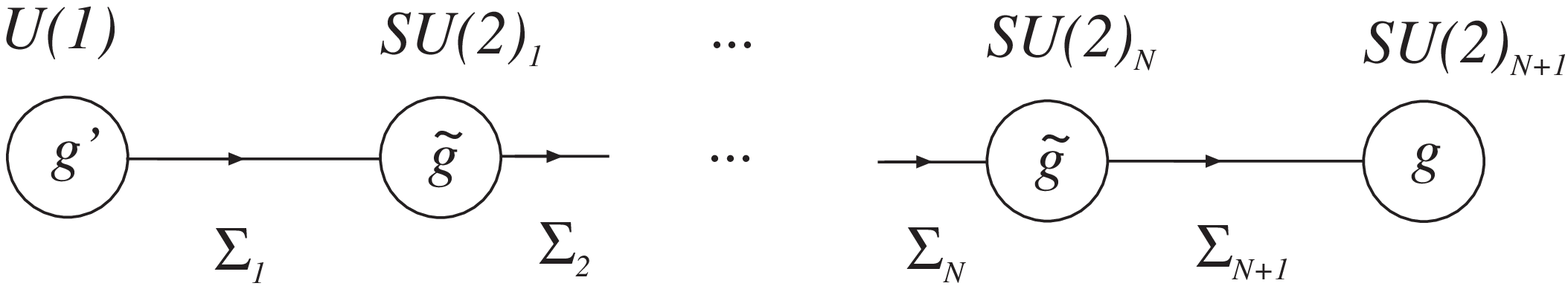,width=0.9\textwidth}
{Moose diagram for $U(1)\times {[SU(2)]}^N\times SU(2)_{N+1}$ model.
\label{fig:MooseN}}

In section \ref{sec:model1} we saw that the choice of parameters
which produced the greatest postponement of unitarity led to
a model where the 
standard model $SU(2)_L$ and $U(1)_Y$ gauge groups
were separated in theory space
by an extra intervening $SU(2)$.  Following the extra-dimensional
analogue further, we now extend this to a series of intervening $SU(2)$'s,
all with the same coupling and all vevs chosen to be the same.
The moose diagram for this theory is shown in Fig.~\ref{fig:MooseN}.

The Lagrange density for this model is
\begin{equation}
{\cal L}\ =\ -{1\over4}B^{\mu\nu}B_{\mu\nu}-{1\over4}\sum_{i=1}^{N+1}W_i^{a\,\mu\nu}W^a_{i\,\mu\nu}
+{f^2\over4}\sum_{i=1}^{N+1}
{\rm tr}\Big[D^\mu\Sigma_i(D_\mu\Sigma_i)^\dagger\Big]\ ,
\label{eq:lagrangeN}
\end{equation}
and the nonlinear sigma model fields can be parameterized by
\begin{equation}
\Sigma_i\ =\ e^{2i\pi_i^aT^a/f}\ .
\label{eq:sigmasN}
\end{equation}
The $\Sigma_i$'s are coupled to the gauge fields
by the covariant derivatives
\begin{eqnarray}
D_\mu\Sigma_1&=& \partial_\mu\Sigma_1-ig^\prime T^3B_\mu\Sigma_1+i\tilde{g}\Sigma_1T^aW^a_{1\mu}\ ,\nonumber\\
D_\mu\Sigma_k&=& \partial_\mu\Sigma_k-i\tilde{g} T^aW^a_{k-1\mu}\Sigma_k+i\tilde{g}
\Sigma_kT^aW^a_{k\mu}\ ,\qquad\qquad(k\ne1,N+1)\nonumber\\
D_\mu\Sigma_{N+1}&=& \partial_\mu\Sigma_{N+1}-i\tilde{g} T^aW^a_{N\mu}\Sigma_{N+1}+ig
\Sigma_{N+1}T^aW^a_{N+1\mu}\ .
\label{eq:covdivN}
\end{eqnarray}
As in the previous model, the sigma fields can be removed in unitary gauge,
giving a mass to the $W$'s and $Z$'s.  We can then expand the 
charged fields in terms of the mass eigenstates
\begin{equation}
W_j^\pm\ =\ \sum_{k=1}^{N}a_{jk}\,W_k^{\prime\pm}+a_{j(N+1)}\,W^{\pm}\ ,
\label{eq:WsN}
\end{equation}
and similarly for the neutral fields
\begin{eqnarray}
B&=& b_{00}\,A+\sum_{k=1}^{N}b_{0k}\,Z_k^\prime+b_{0(N+1)}\,Z\nonumber\\
W_j^3&=& b_{j0}\,A+\sum_{k=1}^{N}b_{jk}\,Z_k^\prime+b_{j(N+1)}\,Z\ ,
\label{eq:ZsN}
\end{eqnarray}
where the photon $A$ is exactly massless as required. 

We give the general solution for the diagonalization of these mass matrices
in
Appendix~\ref{sec:solutions2}.
For this model there are four independent parameters,
$g^\prime,\,\tilde{g},\,g,\,f$, which can be fixed by $e$,
$m_W$, $m_Z$, and the mass of the lightest $W^\prime$, $m_{W_1^\prime}$.
If we assume that $\tilde{g}\gg g, g^\prime$, and letting
$\lambda^2=g^2/\tilde{g}^2$ and $\lambda^{\prime2}=g^{\prime2}/\tilde{g}^2$, 
we obtain the masses 
\begin{eqnarray}
m_W^2&=& {g^2f^2\over4(N+1)}\biggl(1+{\cal O}
(\lambda^2)\biggr)\nonumber\\
m_Z^2&=& {(g^2+g^{\prime2})f^2\over4(N+1)}
\biggl(1+{\cal O}
(\lambda^2)\biggr)\nonumber\\
m_{W_k^\prime}^2&=&\tilde{g}^2f^2\left(\sin{k\pi\over2(N+1)}\right)^2
+2m_W^2\left(\cos{k\pi\over2(N+1)}\right)^2\biggl(1+{\cal O}
(\lambda^2)\biggr)\nonumber\\
m_{Z_k^\prime}^2&=&\tilde{g}^2f^2\left(\sin{k\pi\over2(N+1)}\right)^2
+2m_Z^2\left(\cos{k\pi\over2(N+1)}\right)^2\biggl(1+{\cal O}
(\lambda^2)\biggr)\ .
\label{eq:massesN}
\end{eqnarray}
It is easy to check that for $N=1$ this gives 
$m_{Z_1^\prime}^2- m_{W_1^\prime}^2\approx m_Z^2-m_W^2$,
and $\tilde{g}=(m_{W_1^\prime}/2m_W)g$, up to corrections of order 
$m_W^2/m_{W_1^\prime}^2$, as found in the previous section.

The scattering of longitudinal $W$'s is easily generalized from the
last section. The amplitude for
$W_L^+W_L^-\rightarrow W_L^+W_L^-$ is given by
\begin{eqnarray}
{\cal A}&=& {1\over m_W^4}\Biggl\{\Biggl(\sum_{i=1}^{N+1}g_i^2a_{i(N+1)}^4\Biggr)
\Bigl[p^2E^2(-2+6\cos{\theta})+E^4(-1+\cos^2{\theta})\Bigr]\nonumber\\
&&\qquad+
\Bigl[-p^2\cos{\theta}(p^2-3E^2)^2\Bigr]
\Biggl[{e^2\over s} 
+ \sum_{k=1}^{N+1}{\left(\sum_{i=1}^{N+1}g_i\,b_{ik}\,a_{i(N+1)}^2\right)^2\over 
s-m_{Z_k^{\prime}}^2}
\Biggr]
\nonumber\\
&&\qquad-
\Bigl[4E^2\Bigl(p^2+(E^2-2p^2)\cos{\theta}\Bigr)^2
+2p^2(1+\cos{\theta})\Bigl(2E^2-p^2-E^2\cos{\theta})^2
\Bigr]\nonumber\\
&&\qquad\qquad\times
\Biggl[{e^2\over t} 
+ \sum_{k=1}^{N+1}{\left(\sum_{i=1}^{N+1}g_i\,b_{ik}\,a_{i(N+1)}^2\right)^2\over 
t-m_{Z_k^{\prime}}^2}
\Biggr]\Biggr\}\ .
\label{eq:ampN}
\end{eqnarray}
In this formula we have defined $g_i=\tilde{g}$ for $i\ne N+1$ and 
$g_{N+1}=g$,  
and we have identified $Z_{N+1}^\prime\equiv Z$ in order to make it more
compact.  Then the coefficient of the leading $E^2/m_W^2$ term,
defined by Eq.~(\ref{eq:amp1he}) is
\begin{equation}
R\ =\ 4\left(\sum_{i=1}^{N+1}g_i^2a_{i(N+1)}^4\right)
-3\Biggl[
\sum_{k=1}^{N+1}\left(\sum_{i=1}^{N+1}g_i\,b_{ik}\,a_{i(N+1)}^2\right)^2{m_{Z_k^{\prime}}^2\over m_W^2}
\Biggr]\ .
\label{eq:RN}
\end{equation}
In Appendix~\ref{sec:solutions2} we obtain for this model
\begin{eqnarray}
R&=& {g^2\over(N+1)^2} \ +\ {\cal O}(\lambda^2)\nonumber\\
&=& {R(\,SM\,)\over(N+1)^2}\ +\ {\cal O}(\lambda^2)\ .
\end{eqnarray}
where the corrections also fall off as
$(N+1)^{-2}$.  As expected, this agrees with
the results of the previous section for $N=1$.

\EPSFIGURE[t]{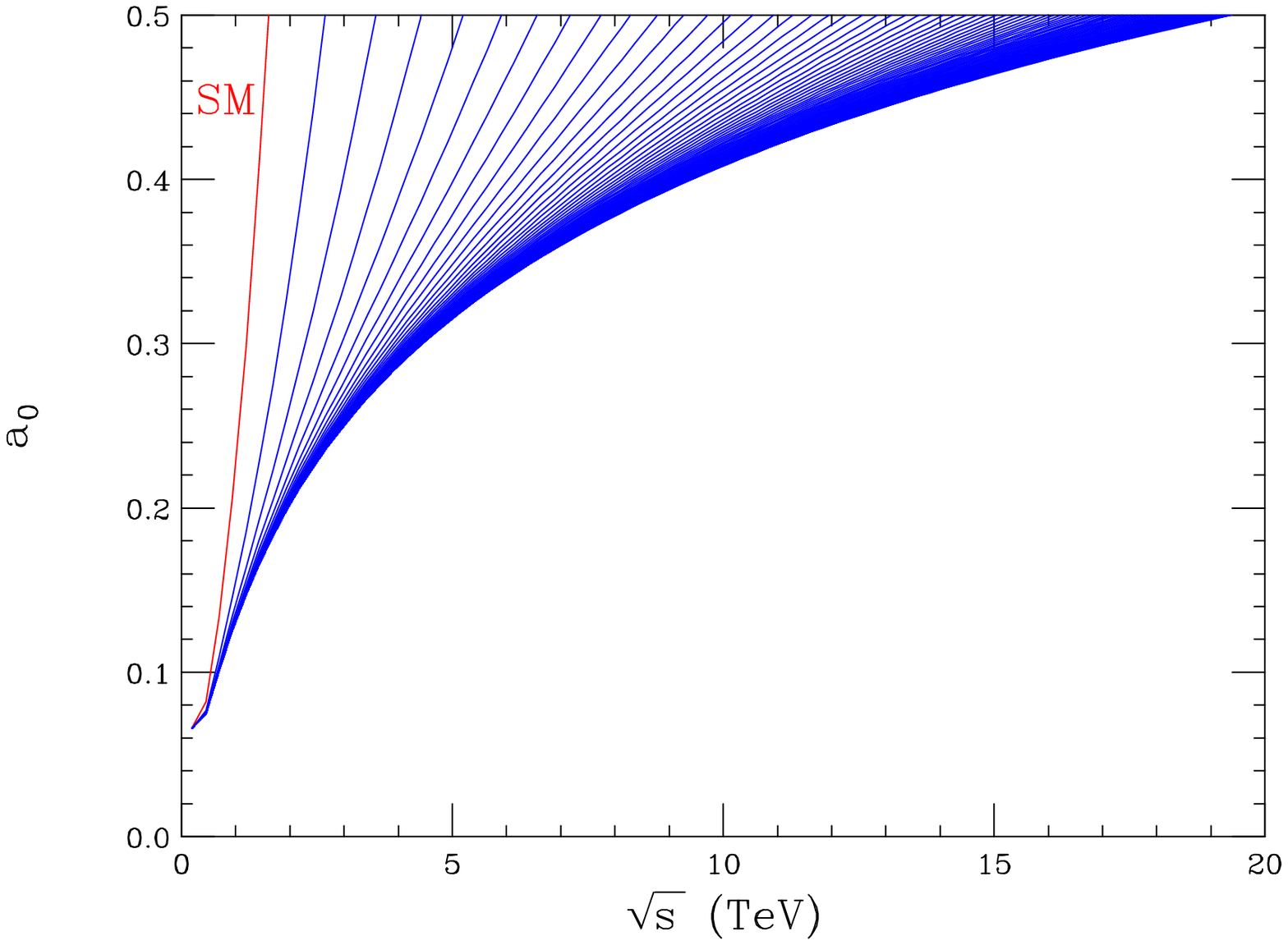,width=0.6\textwidth}
{
The $J=0$ partial wave amplitude as a function of $\sqrt{s}$
for the standard model without a Higgs boson
(red) and the $U(1)\times {[SU(2)]}^N\times SU(2)_{N+1}$ model
(blue) for $N=1$ to 100 with $m_{W_1^\prime}=500$ GeV.
\label{fig:a0N}}

In Fig.~\ref{fig:a0N} we plot the $J=0$ partial wave amplitude
as a function of $\sqrt{s}$
for both the standard model without a Higgs boson and in the
$U(1)\times {[SU(2)]}^N\times SU(2)_{N+1}$
model with $m_{W_1^\prime}=500$ GeV for $N=1$ to 100.
For large $N$ in this model the unitarity violation is delayed to an energy
of about $\sqrt{s}=19$ TeV.  Thus, we may expect that the effective theory
with a KK tower of vector bosons should be reliable up to about 
this scale.\footnote{
A coupled-channel analysis, as considered in Ref.~\cite{sekhar}, would 
give a lower energy scale for unitarity violation.}
At high energies and large $N$, the partial wave amplitude asymptotes to
\begin{eqnarray}
a_0&\approx&{1\over32\pi}\left[{s\over4m_W^2}{g^2\over(N+1)^2}+
{4 \tilde{g}^2\over N+1}\left(
\ln{s\over\Lambda^2}-{1\over2}\right)\right]\nonumber\\
&\approx&{1\over32\pi}\left[{s\over4m_W^2}{g^2\over(N+1)^2}+
{4 g^2\over \pi^2}{m_{W^\prime_1}^2\over m_W^2}\left(
\ln{s\over\Lambda^2}-{1\over2}\right)\right]\ ,
\end{eqnarray}
where $\Lambda$ is a scale on the order of a few times $m_{W^\prime_1}$,
the mass of the lightest $W^\prime$.  

\section{The $N+1\rightarrow\infty$ limit}
\label{sec:edim}

We now consider the $N+1\rightarrow\infty$ limit of this model.  
For large $N$ the sites in the moose diagram of Fig.~\ref{fig:MooseN}
play the role of lattice sites in a fifth dimension, and the model
behaves as a latticized  extra-dimensional theory~\cite{deconstruction}.
Taking the lattice spacing to be $a$ and the length of the extra dimension
to be $\pi R$, we can relate these to the parameters in the model by
\begin{equation}
\pi R\ = \ (N+1)a\ =\ {2(N+1)\over \tilde{g}f}\ .
\end{equation}
The five-dimensional gauge coupling in the bulk is related to the
four-dimensional coupling by $g_5^2=a\tilde{g}^2$, which gives
\begin{equation}
{\tilde{g}^2\over N+1}\ = \ {g_5^2\over\pi R}\ .
\end{equation}
Thus, we find that both the coupling $\tilde{g}$ and the parameter $f$ 
scale as $(N+1)^{1/2}$ as $N+1\rightarrow\infty$.
In this limit the 
model becomes a five-dimensional $SU(2)$ gauge theory, where the fifth 
dimension is the line segment, $0\le y\le \pi R$.
In addition there are 
boundary terms at $y=0$ and $y=\pi R$, 
with the boundary conditions at $y=0$ breaking 
the gauge symmetry down to $U(1)$.  
The five dimensional action is
\begin{eqnarray}
{\cal S}&=& \int_0^{\pi R}dy\int d^4x
\left[
-{1\over4g_5^2}W^{a\,MN}W^a_{MN}
-\delta(y-\pi R){1\over4g^2}W^{a\,\mu\nu}W^a_{\mu\nu}\right.\nonumber\\
&&\qquad\qquad\qquad\quad\left.-\delta(y){1\over4g^{\prime2}}W^{3\,\mu\nu}W^3_{\mu\nu}\right]\ ,
\label{eq:5daction}
\end{eqnarray}
where, in this equation, the indices $M,N$ run over the 5 dimensions,
and we impose the Dirichlet Boundary condition, $W^a_\mu=0$, 
at $y=0$ for $a\ne3$.  The $\delta$-function at $y=\pi R$ 
should be interpreted as $\delta(y-\pi R+\epsilon)$ with 
$\epsilon\rightarrow0^+$ and the fields having Neumann boundary conditions,
 $dW^a_\mu/dy=0$, at $y=\pi R$.  The $\delta$-function 
and the field $W^3_\mu$ at $y=0$ should be interpreted similarly.

All of the results found in appendix~\ref{sec:solutions2} have a well-defined
limit as $N+1\rightarrow\infty$.  The tower of massive vector particles
becomes a tower of Kaluza-Klein states given by the decomposition
\begin{equation}
W^\pm_\mu(x,y)\ =\ \sum_{n=0}^\infty C_{(n)}\,\sin\left(m_{W^{\prime}_n}\,y\right)\,
W^{\prime\pm}_{n,\mu}(x)\ ,
\end{equation}
where $C_{(n)}$ is a normalization factor,
 and the masses $m_{W^{\prime}_n}$ for the $W^{\prime\pm}$ states
are the solutions of
\begin{equation}
m_{W^\prime_n}\tan\left({m_{W^\prime_n}\,\pi R}\right)\ =\ {g^2\over g_5^2}\ .
\label{eq:dim5Wmass}
\end{equation}
The lightest of these charged vector mesons is just the standard model
$W^\pm\equiv W^{\prime\pm}_0$, with mass $m_W\approx g/(g_5\sqrt{\pi R})$.

Similarly the neutral vector bosons can be expanded in Kaluza-Klein
states
\begin{equation}
W^3_\mu(x,y)\ =\ D_{(\gamma)}\, A_\mu(x)\,+\,
\sum_{n=0}^\infty D_{(n)}\,\sin\left(m_{Z^\prime_n}\,y+\phi_{n}\right)\,
Z^{\prime}_{n,\mu}(x)\ ,\label{eq:W3KK}
\end{equation}
where $D_{(\gamma,n)}$ is a normalization factor, the masses $m_{Z^\prime_n}$
of the $Z^\prime$ states are the solutions of
\begin{equation}
\left(m_{Z^\prime_n}^2-{g^2g^{\prime2}\over g_5^4}\right)
\tan\left({m_{Z_n^\prime}\,\pi R}\right)\ =\ {(g^2+g^{\prime2})\over g_5^2}\,m_{Z^\prime_n}\ ,\label{eq:dim5Zmass}
\end{equation}
and the phase constant $\phi_{n}$ satisfies
\begin{equation}
m_{Z^\prime_n}
\tan{\phi_n}\ =\ -{g^{\prime2}\over g_5^2}\ .\label{eq:dim5phase}
\end{equation}
In the Kaluza-Klein expansion (\ref{eq:W3KK}) 
we have separated out the $m_{Z^\prime_n}=0,
\,\phi_n=\pi/2$ 
solution of equations~(\ref{eq:dim5Zmass}) and (\ref{eq:dim5phase}) as 
the massless photon $A$.  The next lightest state
is just the standard model
$Z\equiv Z^{\prime}_0$, with mass 
$m_Z\approx (g^2+g^{\prime2})^{1/2}/(g_5\sqrt{\pi R})$.
The remaining states have masses of approximately
\begin{equation}
m_{W^\prime_n}\ =\ m_{Z^\prime_n}\ =\ {n\over R}\ ,
\end{equation}
up to corrections of ${\cal O}(g^2\,(\pi R/g_5^2))$.

The equations (\ref{eq:dim5Wmass}) and (\ref{eq:dim5Zmass}) which determine the
masses of the Kaluza-Klein towers are the same as those derived in 
Ref.~\cite{Tait}, in a five-dimensional $S_1/Z_2$ orbifold theory 
with non-trivial kinetic terms 
at the orbifold fixed points.\footnote{Our equations actually differ from
Ref.~\cite{Tait} by a factor of 2 in the boundary couplings, which 
just corresponds to a different treatment of the $\delta$-function at the 
boundary.  The $\delta$-function in  
Ref.~\cite{Tait} can be considered as a limiting case of a finite-size brane
which straddles the orbifold fixed point, whereas the $\delta$-function
in eq.~(\ref{eq:5daction}) lies entirely within the interval.}  It is the presence
of these non-trivial kinetic terms at the boundaries
which shift the masses of the
Kaluza-Klein vector bosons so that the lightest states can be identified
as the standard model $W$ and $Z$ bosons (and photon).  
We note that these non-trivial
boundary terms interpolate between the standard Neumann ($\partial_5W=0$)
and Dirichlet ($W=0$) conditions at the boundaries.
For instance, if $g\rightarrow0$,
the solutions to (\ref{eq:dim5Wmass}) correspond to 
Dirichlet boundary conditions at
$y=\pi R$ (with the $W^\pm\equiv W^{\prime\pm}_0$ mode becoming massless and 
decoupling), 
whereas if $g\rightarrow\infty$, they correspond to Neumann boundary 
conditions at $y=\pi R$.
In particular, if we let $g\rightarrow\infty$ and $g^\prime\rightarrow\infty$
we recover the toy $SU(2)\rightarrow U(1)$ model of Ref.~\cite{csaki}.

\section{Coupling to matter fields
and constraints from experiment}
\label{sec:fermions}

In the $U(1)\times {[SU(2)]}^N\times SU(2)_{N+1}$ model the couplings
$g$ and $g^\prime$ are approximately equal to the weak $SU(2)_L$ and
hypercharge $U(1)_Y$ couplings in the standard model, respectively.
Therefore, 
the most obvious choice for incorporating matter fermions in this model
is to couple the left-handed fermions to the $SU(2)_{N+1}$ and 
the left- and right-handed fermions to the $U(1)$ with charges $Y_L$
and $Y_R$, respectively, as in the standard model.  Mass terms for the
fermions can be written in the form
\begin{equation}
{\cal L}_{\rm mass}\ =\ -h f\bar{\psi}_R\Sigma_1\Sigma_2
\cdots\Sigma_{N+1}\psi_L\ +\ {\rm h.c.}\ ,
\label{eq:mass}
\end{equation}
where $h$ is the appropriate Yukawa coupling, and $\bar{\psi}_R=(\bar{u}_R,0)$
 for up-type fermions and $\bar{\psi}_R=(0,\bar{d}_R)$ for down-type 
fermions.  Even though the $\bar{u}_R$ and $\bar{d}_R$ 
fields are not connected by an $SU(2)$ symmetry, it is 
still useful to maintain the 2-component notation for the right-handed fields, 
since the $U(1)$ that is coupled to $\Sigma_1$ is treated as a subgroup of a
global $SU(2)$.  Under the $U(1)$ symmetry  
$\bar{\psi}_R$ has a charge $-Y_R=-Q_{EM}=-(T_3+Y_L)$, $\Sigma_1$
has a charge of $T_3$, and $\psi_L$ has a charge of $Y_L$; thus, the mass
term is invariant under the $U(1)$ symmetry, in addition to the chain of 
$SU(2)$ symmetries.
The extra-dimensional 
interpretation of the $N+1\rightarrow\infty$ limit is not particularly
simple for this manner of incorporation of fermions, since the
left-handed fermions would necessarily have non-zero wave functions
on both the $y=0$ and $y=\pi R$ branes, but would have zero wave functions
in the bulk.
In addition, the mass term (\ref{eq:mass}) would involve a nonlocal 
gauge Wilson loop extending between the two branes.  Nevertheless, the 
$N+1\rightarrow\infty$ limit is well-defined.

Incorporating the fermions into the model in this manner, we can now
investigate what are the experimental constraints on it.  We first
consider the direct constraints from producing the heavy gauge bosons at
colliders and then consider the indirect constraints from precision 
electroweak measurements.

The coupling of fermions to gauge bosons can be written
\begin{equation}
{\cal L}_{Int} = g^\prime~\bar\psi \gamma_\mu 
(Y_LP_L+Y_RP_R)\, \psi~B^\mu + 
             g~\bar\psi \gamma_\mu T^a P_L \psi~W_{N+1}^{a\mu}\ ,
\label{eq:FZp}
\end{equation}
where, $P_{R,L}$ are the projection operators, $(1\pm\gamma_5)/2$.
By expressing $B^\mu$ and $W_{N+1}^{a\mu}$ in terms of the mass
eigenstates using equations (\ref{eq:Ws}) and (\ref{eq:Zs}), we find that 
compared to the $W$ and $Z$ couplings to fermions, the $W^\prime_k$ 
and $Z^\prime_k$ couplings are down by factors of ${\cal O}(\lambda)\approx
{\cal O}(m_W/m_{W^\prime_1})$. 
Due to this suppression, the direct production 
mass bounds at a collider are significantly weakened. The most 
significant bounds on the $W^\prime_1$ and $Z^\prime_1$ masses come from 
the Tevatron and LEP II, respectively. The derivation of a detailed bound 
is beyond the scope of this work and we content ourselves with obtaining 
the following estimates.

The Tevatron (CDF) limit on a $W^\prime$ that couples with Standard Model
strength is presented in Fig.~2 of Ref.~\cite{Affolder:2001gr}. In our case,
the ratio 
$\sigma(q\bar q\rightarrow W^\prime_1 \rightarrow\ell\nu)/\sigma(q\bar q\rightarrow W\rightarrow\ell\nu)$ 
is suppressed by an additional factor of $|a_{(N+1)1}|^2\approx 
k(m_W^2/m_{W^\prime_1}^2)$ to 
that shown 
in the plot, where $k=1$ for $N=1$ and $k=2$ for $N\rightarrow\infty$.
By rescaling the cross sections shown in that figure, we estimate 
that the corresponding limits in our case would be about 
$m_{W_1^\prime} \gtrsim 500$ GeV for $N\rightarrow\infty$,
with the limit for $N=1$ weaker by about 50 GeV.
We should note that in our models,
the $W^\prime_1$ also decays to $WZ$ with roughly the same coupling 
strength as it does to fermions; this may alter the analysis further.

The LEP II bound on new four fermion contact interactions are presented (for 
the case of strong coupling) in Ref.~\cite{:2002mc} by making fits to
$\sigma(e^+e^-\rightarrow f\bar f)$. This can be translated to a bound on
$m_{Z^\prime_1}$ since a heavy $Z^\prime$ effectively induces a four fermion 
contact interaction. Extracting the relevant contact interactions induced
in our model, and comparing to the results of the LEP II analysis, 
we estimate that the mass bound is about $m_{Z_1^\prime} \gtrsim 480$ GeV for
$N\rightarrow\infty$ and $m_{Z_1^\prime} \gtrsim 400$ GeV for $N=1$.

The indirect influence of new physics on electroweak processes 
can be fully parametrized in terms of the oblique parameters $S$, $T$, 
and $U$~\cite{peskin}, as long as the light fermions do not couple directly to
the new, heavy gauge bosons. In our model such coupling 
does occur, but it is suppressed by a factor $\lambda$ in each vertex, in 
addition to the $\lambda^2$ suppression 
that comes from the large mass in the gauge boson propagators at low energies. 
Therefore, the dominant electroweak corrections in our model,
which are ${\cal O}(\lambda^2)$, can be obtained purely 
from the couplings of fermions to the standard model $W$ and $Z$ bosons, 
both at low energies and at the $Z$-pole, without considering the direct
fermion coupling to the heavier mass eigenstates.  
Deviations from standard model relations in the $W$ and $Z$ couplings can
then be parametrized in terms of $S$, $T$, and $U$. 
Following Ref.~\cite{burgess}, we have expressed the charged-current
and neutral-current interactions for the physical $W$ and $Z$ bosons, 
obtained from eq.~(\ref{eq:FZp}), in terms
of the electric charge $e$, the $Z$ boson mass $m_Z$, and the Fermi constant
$G_F$, plus new physics contributions.  Comparing to the general expressions
in Ref.~\cite{burgess}, we obtain
\begin{eqnarray}
\alpha S &=& \frac{2N(N+2)}{3(N+1)}\frac{\lambda^2\lambda'^2}{\lambda^2+\lambda'^2} \nonumber\\
&=&\frac{8N(N+2)}{3}
\left(\sin\frac{\pi}{2(N+1)}\right)^2\frac{m_W^2}{m_{W^\prime_1}^2}\sin^2\theta_W \ ,\nonumber\\
T &=& 0 \ ,\nonumber\\
U &=& 0 \ .\label{eq:STU}
\end{eqnarray}
The vanishing of $T$ and $U$, at this order, is a consequence of the coupling 
of $B_\mu$ as the $T^3$ component of a global $SU(2)$, a choice which 
preserves an approximate custodial 
symmetry. For the minimal ($N=1$) model we have 
\begin{equation}
\alpha S = 4\frac{m_W^2}{m_{W^\prime_1}^2}\sin^2\theta_W\ ,\label{eq:S,N=1}
\end{equation}
whereas a true fifth dimension ($N\to\infty$) gives
\begin{equation}
\alpha S = \frac{2\pi^2}{3}\frac{m_W^2}{m_{W^\prime_1}^2}\sin^2\theta_W\ .\label{eq:S,N=infty}
\end{equation}

\EPSFIGURE[t]{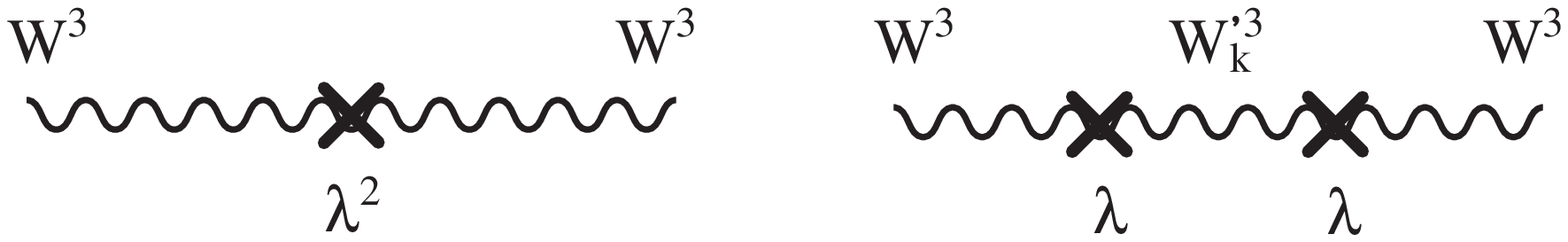,width=0.6\textwidth}
{Contributions to $\Pi_{33}(q^2)$ from mixing of the gauge bosons
at order $\lambda^2$.
\label{fig:Sdiagram}}

It is also possible to obtain these corrections by directly applying the
definition of $S$, $T$, and $U$ in terms of gauge-eigenstate vacuum 
polarizations, as given in Ref.~\cite{peskin}.  Since
this offers some insight into the origin of the corrections, we give an
overview of this approach here.
At tree level the fermions couple only to the gauge eigenstates 
$W^a_{N+1}$ and $B$, so all interactions can be calculated using the
full propagators of these
gauge eigenstates.  In this sense the re-diagonalization of the 
gauge eigenstates into mass eigenstates can be considered purely ``oblique''; 
{\it i.e.,} it only modifies the 
gauge-eigenstate propagators and does not give any 
``direct'' vertex corrections.
We can calculate these tree-level oblique corrections as a 
perturbation expansion in $\lambda$ and $\lambda^\prime$, 
in exact analogy to the
loop-level perturbative corrections that occur in the standard model.
The parameter $S$ is defined by
\begin{equation}
\alpha S = 4e^2[\Pi^\prime_{33}(0)-\Pi^\prime_{3Q}(0)]\ ,\label{eq:Sdef}
\end{equation}
where $\Pi^\prime_{XY}(q^2)$ is related to the vacuum-polarization amplitude 
between the gauge-eigenstate currents $J_X$ and $J_Y$ by 
$\Pi_{XY}(q^2)=\Pi_{XY}(0)+q^2\Pi^\prime_{XY}(q^2)$.
At zeroeth order in $\lambda$ and $\lambda^\prime$, the standard model $W^\pm$,
 $Z$, and photon are obtained from linear combinations of the ``brane''
gauge bosons $W^a_{N+1}$ and $B$. 
Meanwhile, the remaining ``bulk'' gauge eigenstates $W^a_i$ 
are diagonalized into mass eigenstates $W^{\prime a}_j$ by
$W^a_i=\sum a_{ij}W^{\prime a}_j$ for $i,j=1$ to $N$.  The masses
of these states, $m_{W^\prime_j}$, and the mixings, $a_{ij}$, 
can be found from the $\lambda=\lambda^\prime=0$ 
limit of the masses and mixings of the heavy gauge boson states, given
in appendix \ref{sec:solutions2}.  Contributions to the $\Pi$'s
then arise at ${\cal O}(\lambda^2)$ via the perturbative 
mixing between the ``bulk'' and ``brane'' gauge bosons.
For example, the Feynman diagrams for the contribution to $\Pi_{33}$
are shown in Fig.~\ref{fig:Sdiagram}, where the mass insertions come from
the terms in the mass matrix of eq.~(\ref{eq:Zmassmatrix}) 
which are proportional to $\lambda^2$ and $\lambda$.
Noting that the mass insertion proportional to $\lambda$ only has
couplings between the $W^3_{N+1}$ and $W^3_N$ gauge eigenstates, we
obtain
\begin{eqnarray}
ig^2\Pi_{33}(q^2)&=&\left\{\frac{i\lambda^2\tilde{g}^2f^2}{4}\ +\ \sum_{k=1}^N 
\left(\frac{-i\lambda\tilde{g}^2 f^2}{4}
\right)
\left(\frac{-i\,|a_{Nk}|^2}{q^2-m^{2}_{W^\prime_k}}\right) 
\left(\frac{-i\lambda\tilde{g}^2 f^2}{4}\right)
\right\}\,.\label{eq:Pi_33}
\end{eqnarray}
where we have expanded the $\langle W_N^3 W_N^3\rangle$ propagator in terms
of the mass eigenstate propagators. In calculating $\Pi_{3Q}$ we
also need $\langle W_N^3 W_1^3\rangle$, whose expansion in terms of the 
mass eigenstate propagators involves the coefficients $a_{Nk}a_{1k}^*$. 
This leads to the following expression:
\begin{equation}
\Pi'_{33}(0)-\Pi'_{3Q}(0)=\frac{\tilde{g}^2 f^4}{16}\left[-\sum_{k=1}^N 
\frac{|a_{Nk}|^2}{m_{W'_k}^4}\ + \sum_{k=1}^N \frac{a_{Nk}\,(a_{1k}^*+a_{Nk}^*)}{m_{W'_k}^4}\right]\,.
\label{eq:Pi_33-Pi_3Q}
\end{equation}
Thus, we find a nonzero contribution to $S$ from the presence of the
$a_{Nk}a_{1k}^*$ terms, which do not cancel between
$\Pi'_{33}(0)$ and $\Pi'_{3Q}(0)$.  Using the formulae in 
Appendix~\ref{sec:solutions2} and some algebraic identities, we then
recover the results of Eq.(\ref{eq:STU}).\footnote{In this approach we
note that any other oblique parameters which might distinguish between
low-energy and $Z$-pole processes, such as those considered in
Ref.~\cite{Maksymyk}, have an additional $\lambda^2$ suppression 
factor, as expected from preceding arguments.}

Recent experimental constraints on $S$ and $T$ have been compiled in 
Ref.~\cite{PDG}, where the limits are given as a function of the
Higgs boson mass.  In principle, its contributions must be 
subtracted from the above $S$ and $T$ parameters, since there is no 
Higgs boson in our model.  However, given that the dependence on 
$m_H$ is not too large, we can still obtain an estimate of how these 
constraints impact our model.  For $m_H=600$ GeV
with the constraint $S\geq 0$, and using Bayesian statistics,  the limit
on $S$ is $S\leq 0.14$.  This result corresponds to 
$m_{W^\prime_1} \gtrsim 2.3$ TeV ($N=1$) and  $m_{W^\prime_1} \gtrsim 3$ TeV ($N\to\infty$).
Unfortunately, for models in which the first $W^\prime$ mass is so large, 
unitarity will be violated even before the scale of the first $W^\prime$ is
reached.  Therefore, it appears that the method used in this section to
incorporate matter fermions into the model is not viable.

\section{Conclusions}
\label{sec:conclusions}

In this paper we have investigated a class of theory-space models,
which consist of a chain of $SU(2)$ gauge theories and one $U(1)$ gauge
theory linked together by non-linear sigma fields, and  
which contain no physical Higgs scalar.  By allowing the couplings of the 
gauge fields at the first and last site to differ from the others, 
we can obtain the $W^\pm$ and $Z$ boson masses for the first charged
and neutral states in a tower of massive vector bosons (in addition to 
a massless state, the photon).  Furthermore, we have found that 
the unitarity violation in $W^+_LW^-_L$ scattering, which one  
finds in the standard model without a physical Higgs boson, is delayed 
by the exchange of the tower of heavy vector boson states~\cite{sekhar}.  

In the $N+1\rightarrow\infty$ limit of these models, 
we find that the coefficient of the
leading $E^2/m_W^2$ term in the scattering amplitude vanishes.  In this
limit, these models become an $SU(2)$ gauge theory defined on a 
fifth-dimensional line segment, with the $SU(2)$ broken explicitly to 
$U(1)$ at one boundary, and with nontrivial kinetic terms at both boundaries.  
The effective four-dimensional gauge sector reduces to the standard model 
gauge sector at low energies and can be made unitary to scales of 10 TeV or 
more.  At the gauge-sector level, this model is simpler than those
proposed previously~\cite{csaki,nomura,barbieri}, all of which were based on 
the gauge group $SU(2)\times SU(2)\times U(1)$.

We also considered a particular implementation of matter fermions coupled
to these models.  In this implementation the fermions only couple directly
to the first and last gauge fields in the chain.  This naturally gives
standard-model-like couplings of the fermions to the $W^\pm$ and $Z$ bosons,
while the couplings to the heavier vector bosons are suppressed by
ratios of the $W$ mass to the mass of the next higher $W^\prime$ state.
In this way the bounds on the $W^\prime$ and $Z^\prime$ masses are
lowered compared to models where the $W^\prime$ and $Z^\prime$
have standard-model-like couplings.

We found that the most serious constraint on this particular implementation
of fermions comes from electroweak precision measurements.  Due to the
mixing between the vector bosons, there is a small deviation from
the standard model prediction, which can be parameterized (to first
order in $m_W^2/m_{W^\prime_1}^2$) in terms of the oblique correction
parameters $S$, $T$, and $U$.  In our models we find $T=U=0$ occurs
naturally, because of an approximate custodial $SU(2)$.  This occurs
without the need to introduce an additional $SU(2)$ gauge field in the
bulk as in refs.~\cite{csaki,nomura,barbieri}; in our models the 
already-present $SU(2)$ gauge symmetry at adjacent positions in the 
latticized fifth dimension provides the necessary custodial symmetry.
However, 
there is a reasonably-sized contribution to $S$.  In fact, the 
electroweak precision constraints on $S$ probably 
force the scale of $m_{W^\prime_1}$ to be too large to fix the 
unitarity problem which initially motivated our investigation.

This non-zero contribution to $S$ has also been found in
other attempts at Higgsless models of electroweak-symmetry 
breaking~\cite{barbieri}.
It is interesting that these models, which appear similar to 
technicolor models in many ways, 
should have the same difficulty with the $S$
parameter that plagued the simplest versions of technicolor~\cite{peskin}.
However, these models offer a new approach to this problem, with the
possibility to find a mechanism to make $S$ ``naturally'' small.  

Finally, we note that the implementation of the matter fields presented
in this paper does not have a simple extra-dimensional 
interpretation in the $N+1\rightarrow\infty$ limit.
The difficulty arises
because the left-handed fields have non-zero hypercharge ($Y_L$), and therefore
must couple at both ends of the fifth-dimensional interval.  
Even if we let the fermions extend into the bulk (thereby allowing a 
local mass term), we find that the fermions at all positions along the 
line segment
must couple to $W^3(x,y=0)$, due to the non-zero $Y_L$.
Although this may be considered purely a question of aesthetics, since the
present models are no less well-defined than a five-dimensional theory, it is 
still worthwhile to consider this issue and
think of other generalizations.  


\vskip .2 cm

\section*{Acknowledgments}

This work was supported by the US National
Science Foundation under grants PHY-0070443
and PHY-0244789. We would like to thank Sekhar Chivukula and
Tim Tait for useful discussions.

\appendix

\section{Solutions for the $U(1)\times SU(2)_1\times SU(2)_2$ Model}
\label{sec:solutions1}

In this model there are five independent parameters: $g^\prime,\,
\tilde{g},\,g,\,f_1,\,f_2$.  We can rewrite these as functions
of the standard model parameters $e,\,m_W,\,m_Z$ and
the masses of the heavy vector bosons, $m_{W^\prime},\,
m_{Z^\prime}$.  We obtain
\begin{eqnarray}
g^{\prime2}&=&{e^2m_Z^2m_{Z^\prime}^2\over m_W^2m_{W^\prime}^2}\ ,\nonumber\\
\tilde{g}^{2}&=&g^{\prime2}\left[
{(m_W^2+m_{W^\prime}^2)(m_Z^2+m_{Z^\prime}^2-m_W^2-m_{W^\prime}^2)
+m_W^2m_{W^\prime}^2-m_Z^2m_{Z^\prime}^2\over
(m_Z^2+m_{Z^\prime}^2-m_W^2-m_{W^\prime}^2)^2}\right]\ ,\nonumber\\
g^{2}&=&{g^{\prime2}m_W^2m_{W^\prime}^2}\left[
{(m_W^2+m_{W^\prime}^2)(m_Z^2+m_{Z^\prime}^2-m_W^2-m_{W^\prime}^2)
+m_W^2m_{W^\prime}^2-m_Z^2m_{Z^\prime}^2\over
(m_{Z}^2-m_{W}^2)(m_{Z^\prime}^2-m_{W^\prime}^2)
(m_{Z^\prime}^2-m_{W}^2)(m_{W^\prime}^2-m_{Z}^2)}\right]\ ,\nonumber\\
f_1^{2}&=&{4\over g^{\prime2}}
(m_Z^2+m_{Z^\prime}^2-m_W^2-m_{W^\prime}^2)\ ,\nonumber\\
f_2^{2}&=&{16m_W^2m_{W^\prime}^2\over\tilde{g}^2g^2f_1^2}\ .
\label{params}
\end{eqnarray}
Note these equations imply the relationship
$m_Z^\prime>m_W^\prime$.

The charged boson mixing matrix, defined in Eq.~(\ref{eq:Ws}),
is given by $a_{11}=a_{22}=\cos{\phi}$ and 
$a_{12}=-a_{21}=\sin{\phi}$ with 
\begin{eqnarray}
\cos{\phi}& =& \left[{m_{W^\prime}^2(m_{W^\prime}^2-m_{Z}^2)
(m_{Z^\prime}^2-m_{W^\prime}^2)\over
m_{W^\prime}^2(m_{W^\prime}^2-m_{Z}^2)
(m_{Z^\prime}^2-m_{W^\prime}^2)+m_{W}^2(m_{Z^\prime}^2-m_{W}^2)
(m_{Z}^2-m_{W}^2)}
\right]^{1/2}\ ,\nonumber\\
\sin{\phi}& =& \left[{m_{W}^2(m_{Z^\prime}^2-m_{W}^2)
(m_{Z}^2-m_{W}^2)\over
m_{W^\prime}^2(m_{W^\prime}^2-m_{Z}^2)
(m_{Z^\prime}^2-m_{W^\prime}^2)+m_{W}^2(m_{Z^\prime}^2-m_{W}^2)
(m_{Z}^2-m_{W}^2)}
\right]^{1/2}\ .
\label{eq:Wmix}
\end{eqnarray}
The neutral boson mixing matrix, defined in Eq.~(\ref{eq:Zs}),
is given by
\begin{equation}
b_{00}\ =\ e/g^\prime\,,\qquad\ b_{10}\ =\ e/\tilde{g}\,,
\qquad\ b_{20}\ =\ e/g\,,
\end{equation}
and
\begin{eqnarray}
b_{01}&=&-\left[{
(m_{Z^\prime}^2-m_{W}^2)(m_{Z^\prime}^2-m_{W^\prime}^2)
\over m_{Z^\prime}^2(m_{Z^\prime}^2-m_{Z}^2)}
\right]^{1/2}\ ,\nonumber\\
b_{11}&=&\left[{
(m_{Z^\prime}^2-m_{W}^2)(m_{Z^\prime}^2-m_{W^\prime}^2)
\over m_{Z^\prime}^2(m_{Z^\prime}^2-m_{Z}^2)
\Bigl[(m_W^2+m_{W^\prime}^2)(m_Z^2+m_{Z^\prime}^2-m_W^2-m_{W^\prime}^2)
+m_W^2m_{W^\prime}^2-m_Z^2m_{Z^\prime}^2\Bigr]
}
\right]^{1/2}\nonumber\\
&&\times\biggl(m_{W^\prime}^2+m_{W}^2-m_Z^2\biggr)\ ,\nonumber\\
b_{21}&=&-\left[{m_W^2m_{W^\prime}^2
(m_{Z}^2-m_{W}^2)(m_{W^\prime}^2-m_{Z}^2)
\over m_{Z^\prime}^2(m_{Z^\prime}^2-m_{Z}^2)
\Bigl[(m_W^2+m_{W^\prime}^2)(m_Z^2+m_{Z^\prime}^2-m_W^2-m_{W^\prime}^2)
+m_W^2m_{W^\prime}^2-m_Z^2m_{Z^\prime}^2\Bigr]
}
\right]^{1/2}\ ,\nonumber\\
b_{02}&=&-\left[{
(m_{Z}^2-m_{W}^2)(m_{W^\prime}^2-m_{Z}^2)
\over m_{Z}^2(m_{Z^\prime}^2-m_{Z}^2)}
\right]^{1/2}\ ,\\
b_{12}&=&\left[{
(m_{Z}^2-m_{W}^2)(m_{W^\prime}^2-m_{Z}^2)
\over m_{Z}^2(m_{Z^\prime}^2-m_{Z}^2)
\Bigl[(m_W^2+m_{W^\prime}^2)(m_Z^2+m_{Z^\prime}^2-m_W^2-m_{W^\prime}^2)
+m_W^2m_{W^\prime}^2-m_Z^2m_{Z^\prime}^2\Bigr]
}
\right]^{1/2}\nonumber\\
&&\times\biggl(m_{W^\prime}^2+m_{W}^2-m_{Z^\prime}^2\biggr)\ ,\nonumber\\
b_{22}&=&\left[{m_W^2m_{W^\prime}^2
(m_{Z^\prime}^2-m_{W}^2)(m_{Z^\prime}^2-m_{W^\prime}^2)
\over m_{Z}^2(m_{Z^\prime}^2-m_{Z}^2)
\Bigl[(m_W^2+m_{W^\prime}^2)(m_Z^2+m_{Z^\prime}^2-m_W^2-m_{W^\prime}^2)
+m_W^2m_{W^\prime}^2-m_Z^2m_{Z^\prime}^2\Bigr]
}
\right]^{1/2}\ .\nonumber
\label{eq:bijs}
\end{eqnarray}

\section{Solutions for the $U(1)\times {[SU(2)]}^N\times SU(2)_{N+1}$ Model}
\label{sec:solutions2}

The mass matrix for the charged bosons is
\begin{equation}
M_W^2\ =\ {\tilde{g}^2f^2\over4}
\left[ \begin{array}{ccccccc}
2  & -1 &  0 & \cdots  &&&\\
-1 &  2 & -1 & \cdots& &&\\
0  & -1 &  2 &\cdots  &&&\\
\vdots &\vdots & \vdots &\ddots &\vdots&\vdots&\vdots\\
    &   &  &\cdots&  2 & -1 & 0\\
 &&& \cdots& -1 & 2 &-\lambda\\
&&& \cdots&  0 & -\lambda &\lambda^2 
\end{array}\right]\ ,
\end{equation}
where $\lambda=g/\tilde{g}$.
Note that the sequence of $(-1\ 2\ -1)$ in each row of this matrix
(except the first and the last two)
acts as a discrete second derivative, whose eigenfunction is a
sine function.  Thus, we obtain as eigenvectors for this
matrix 
\begin{equation}
\psi_{(n)}\ =\ C_{(n)}\left[
\begin{array}{c}
\sin{\omega_{(n)}}\\
\sin{2\omega_{(n)}}\\
\sin{3\omega_{(n)}}\\
\vdots\\
\sin{(N-1)\omega_{(n)}}\\
\sin{N\omega_{(n)}}\\
{1\over\lambda}\sin{(N+1)\omega_{(n)}}\\
\end{array}
\right]\ ,
\end{equation}
where $C_{(n)}$ is a normalization constant and
the eigenvalues are
\begin{equation}
m_{W^\prime_n}^2\ =\ \tilde{g}^2f^2\sin^2{\omega_{(n)}\over2}\ .
\end{equation}
The last row of the eigenvector equation 
$M_W^2\psi_{(n)}=m_{W^\prime_n}^2\psi_{(n)}$
gives the characteristic equation for this system
\begin{equation}
\sin^2{\omega\over2}\sin{(N+1)\omega}\ =\ {\lambda^2\over4}\Bigl[
\sin{(N+1)\omega}-\sin{N\omega}
\Bigr]\ ,
\label{eq:ChiW}
\end{equation}
which has $N+1$ solutions, $\omega_{(n)}$.
Using this equation and trigonometric identities~\cite{GR},
we obtain a simple formula for the normalization constant
\begin{equation}
C_{(n)}\ =\ \left[{N+1\over2}+{\sin{\left[2(N+1)\omega_{(n)}\right]}\over4\sin{\omega_{(n)}}}\right]^{-1/2}
\end{equation}

Solving Eq.~(\ref{eq:ChiW}) perturbatively, and identifying
the standard model $W\equiv W^\prime_{N+1}$, we obtain for the
charged boson masses
\begin{eqnarray}
m_W^2&=&{g^2f^2\over4(N+1)}\left(1-\lambda^2{N(2N+1)\over6(N+1)}+{\cal O}(\lambda^4)\right)\ ,\nonumber\\
m_{W^\prime_n}^2&=&\tilde{g}^2f^2\left(\sin{n\pi\over2(N+1)}\right)^2
+2m_W^2\left(\cos{n\pi\over2(N+1)}\right)^2\biggl(1+{\cal O}
(\lambda^2)\biggr)\ .
\end{eqnarray}
The elements of the charged boson mixing matrix are
\begin{eqnarray}
a_{(N+1)(N+1)}&=&1-\lambda^2{N(2N+1)\over12(N+1)}+{\cal O}(\lambda^4)\ ,\nonumber\\
a_{n(N+1)}&=&\lambda{n\over N+1}+{\cal O}(\lambda^3)\ ,\nonumber\\
a_{(N+1)m}&=&-\lambda\sqrt{2\over N+1}\,{\sin{\pi m N\over N+1}\over 
4\sin^2{\pi m\over2(N+1)}}+{\cal O}(\lambda^3)\ ,\nonumber\\
a_{nm}&=&\sqrt{2\over N+1}\,\sin{\pi nm \over N+1}+{\cal O}(\lambda^2)\ ,
\label{eq:anm}
\end{eqnarray}
where $n$ and $m$ run from 1 to $N$.

The mass matrix for the neutral bosons is
\begin{equation}
M_Z^2\ =\ {\tilde{g}^2f^2\over4}
\left[ \begin{array}{ccccccc}
\lambda^{\prime2}  & -\lambda^\prime &  0 & \cdots  &&&\\
-\lambda^\prime &  2 & -1 & \cdots& &&\\
0  & -1 &  2 &\cdots  &&&\\
\vdots &\vdots & \vdots &\ddots &\vdots&\vdots&\vdots\\
    &   &  &\cdots&  2 & -1 & 0\\
 &&& \cdots& -1 & 2 &-\lambda\\
&&& \cdots&  0 & -\lambda &\lambda^2 
\end{array}\right]\ ,\label{eq:Zmassmatrix}
\end{equation}
where $\lambda^\prime=g^\prime/\tilde{g}$.
The eigenvector equation 
$M_Z^2\chi_{(n)}=m_{Z^\prime_n}^2\chi_{(n)}$
can be solved in a similar manner to the charged mass matrix.
The eigenvectors are
\begin{equation}
\chi_{(n)}\ =\ D_{(n)}\left[
\begin{array}{c}
{1\over\lambda^\prime}\sin{\phi_{(n)}}\\
\sin{[\rho_{(n)}+\phi_{(n)}]}\\
\sin{[2\rho_{(n)}+\phi_{(n)}]}\\
\vdots\\
\sin{[(N-1)\rho_{(n)}+\phi_{(n)}]}\\
\sin{[N\rho_{(n)}+\phi_{(n)}]}\\
{1\over\lambda}\sin{[(N+1)\rho_{(n)}+\phi_{(n)}]}\\
\end{array}
\right]\ ,
\end{equation}
where $D_{(n)}$ is a normalization constant and
the eigenvalues are
\begin{equation}
m_{Z^\prime_n}^2\ =\ \tilde{g}^2f^2\sin^2{\rho_{(n)}\over2}\ .
\end{equation}
The characteristic equation for this system is
\begin{equation}
\sin^2{\rho\over2}\sin{(N+1)\rho}\ =\ {\lambda^2+\lambda^{\prime2}\over4}\Bigl[
\sin{(N+1)\rho}-\sin{N\rho}
\Bigr]+{\lambda^2\lambda^{\prime2}\over4}\sin{N\rho}\ ,
\label{eq:ChiZ}
\end{equation}
which has $N+2$ solutions, $\rho_{(n)}$. The phase constant $\phi_{(n)}$
satisfies 
\begin{equation}
\tan{\phi_{(n)}}\tan{\rho_{(n)}\over2}\ =\ {\lambda^{\prime2}\over\lambda^{\prime2}-2}
\ .
\label{eq:phase}
\end{equation}
Using Eqs.~(\ref{eq:ChiZ}) and (\ref{eq:phase}), we obtain for the
normalization constant
\begin{equation}
D_{(n)}\ =\ \left[{N+1\over2}+{\sin{\left[(N+1)\rho_{(n)}\right]}\cos{\left[(N+1)\rho_{(n)}
+2\phi_{(n)}\right]}\over2\sin{\rho_{(n)}}}\right]^{-1/2}
\end{equation}

There is one trivial solution to Eqs.~(\ref{eq:ChiZ}) and (\ref{eq:phase}),
which corresponds to the photon solution.  
Identifying the photon by $A\equiv Z^\prime_{0}$, we obtain
the solution $\rho_{(0)}=0$, $\phi_{(0)}=\pi/2$.  The photon is massless,
and the electromagnetic coupling is given by
\begin{equation}
{1\over e^2}\ =\ {1\over g^{\prime2}}+{N\over \tilde{g}^{2}}+{1\over g^{2}}\ .
\end{equation}
The photon mixing angles are
\begin{equation}
b_{00}\ =\ e/g^\prime\,,\qquad\ b_{n0}\ =\ e/\tilde{g}\,,
\qquad\ b_{(N+1)0}\ =\ e/g\,,
\end{equation}
with $n=1$ to $N$.

Identifying
the standard model $Z\equiv Z^\prime_{N+1}$, we obtain for the
the remaining neutral boson masses
\begin{eqnarray}
m_Z^2&=&{(g^2+g^{\prime2})f^2\over4(N+1)}\left(1-
(\lambda^2+\lambda^{\prime2}){N(2N+1)\over6(N+1)}
+{N\lambda^2\lambda^{\prime2}\over\lambda^2+\lambda^{\prime2}}
+{\cal O}(\lambda^4)\right)\ ,\nonumber\\
m_{Z^\prime_n}^2&=&\tilde{g}^2f^2\left(\sin{n\pi\over2(N+1)}\right)^2
+2m_Z^2\left(\cos{n\pi\over2(N+1)}\right)^2\biggl(1+{\cal O}
(\lambda^2)\biggr)\ .
\end{eqnarray}
The elements of the charged boson mixing matrix are
\begin{eqnarray}
b_{0(N+1)}&=&{-g^{\prime}\over\sqrt{g^2+g^{\prime2}}}\left[
1-(\lambda^2+\lambda^{\prime2}){N(2N+1)\over12(N+1)}
+{N\lambda^4\over2(\lambda^2+\lambda^{\prime2})}
+{\cal O}(\lambda^4)\right]\ ,\nonumber\\
b_{n(N+1)}
&=&{1\over N+1}\left[
n{\lambda^2\over\sqrt{\lambda^2+\lambda^{\prime2}}}
-(N+1-n){\lambda^{\prime2}\over\sqrt{\lambda^2+\lambda^{\prime2}}}
+{\cal O}(\lambda^3)\right]\ ,\nonumber\\
b_{(N+1)(N+1)}
&=&{g\over\sqrt{g^2+g^{\prime2}}}\left[
1-(\lambda^2+\lambda^{\prime2}){N(2N+1)\over12(N+1)}
+{N\lambda^{\prime4}\over2(\lambda^{2}+\lambda^{\prime2})}
+{\cal O}(\lambda^4)\right]\ ,\nonumber\\
b_{0m}
&=&-\lambda^\prime\sqrt{2\over N+1}\,{\sin{\pi m \over N+1}\over 
4\sin^2{\pi m\over2(N+1)}}+{\cal O}(\lambda^3)\ ,\nonumber\\
b_{nm}
&=&\sqrt{2\over N+1}\,\sin{\pi nm \over N+1}+{\cal O}(\lambda^2)\ ,\nonumber\\
b_{(N+1)m}
&=&-\lambda\sqrt{2\over N+1}\,{\sin{\pi m N \over N+1}\over 
4\sin^2{\pi m\over2(N+1)}}+{\cal O}(\lambda^3)\ ,
\label{eq:bnm}
\end{eqnarray}
where $n$ and $m$ run from 1 to $N$.

Finally, we can use the characteristic equations, (\ref{eq:ChiW}),
(\ref{eq:ChiZ}), and (\ref{eq:phase}), along with 
the orthonormality of the rows of the $Z^\prime$
mixing matrix, to obtain a simple expression for the leading $E^2/m_{W_n^{\prime}}^2$
term in the $W^{\prime+}_nW^{\prime-}_n\rightarrow W^{\prime+}_nW^{\prime-}_n$ scattering amplitude, which is the generalization of $R$ in Eq.~(\ref{eq:RN}).
We find 
\begin{equation}
R_{(n)}\ =\ C_{(n)}^4\left({m_{W^\prime_n}^2\over f^2}\right)
\left[{3\over2}(N+1)+{\sin{[2(N+1)\omega_{(n)}]}\over\sin{\omega_{(n)}}}
+{\sin{[4(N+1)\omega_{(n)}]}\over4\sin{2\omega_{(n)}}}\right]\ .
\end{equation}
It is interesting to note that this quantity is exactly independent of
$g^\prime$, and it falls off as $(N+1)^{-2}$ for large $N$.
Setting $n=N+1$, we obtain the result for $W^+W^-$ 
scattering in this model which, to first non-zero order in $\lambda^2$, is
\begin{equation}
R\ =\ {g^2\over(N+1)^2}\ .
\end{equation}


\end{document}